\documentclass[fleqn,usenatbib]{mnras}
\usepackage{color}
\usepackage[T1]{fontenc}
\usepackage{amsmath}	% Advanced maths commands
\usepackage{amssymb}	% Extra maths symbols
\usepackage{ae,aecompl}
\usepackage{graphicx}
\usepackage{rotating}
\usepackage{lscape}
\usepackage{amstext}
\usepackage{multirow}
\usepackage{hyperref}
\usepackage{caption}

\title[An MCMC Approach to the Structure of the Galactic bulge]
{An MCMC Approach to the Three-dimensional Structure of the Milky Way Bulge 
using OGLE-IV $\delta$ Scuti Stars}
\author[Mami Deka et al.]{Mami Deka$^{1}$\thanks{E-mail:mamideka8@gmail.com}, 
Sukanta Deb$^{1,2}$\thanks{E-mail:sukanta.deb@cottonuniversity.ac.in}, 
Kerdaris Kurbah$^{1}$ \\ 
$^{1}$Department of Physics, Cotton University, Panbazar, Guwahati 781001,
Assam, India \\
$^{2}$Space and Astronomy Research Center, Cotton University, Panbazar, Guwahati 781001, Assam, India \\
}
\date{Received on ; Accepted on }
\pubyear{2021}
\begin{document}
\label{firstpage}
\pagerange{\pageref{firstpage}--\pageref{lastpage}}
\maketitle
\begin{abstract}
We present an analysis of high latitude  $\delta$ Scuti stars 
($\left|b\right|> 1^{\circ}$) in the Galactic bulge region 
($-8^{\circ}.3< l<9^{\circ}.4$) using a clean sample of the photometric data 
of $7,440$ stars recently released by the OGLE-IV project. The geometrical 
parameters of the bulge are determined based on Maximum Likelihood (ML) 
analysis in five-dimensional parameter space. More refined values of these 
parameters as well as their uncertainties are obtained from a fully Bayesian 
Markov Chain Monte Carlo (MCMC) analysis. Approximating the bulge as an 
ellipsoid, the distribution of the number density of stars as a function of 
Galacto-centric distance has been modelled 
using three distribution functions: two Exponential ($\rm E_{1},\rm E_{2}$) 
types and one Gaussian ($\rm G$) type. 
Based on the AIC and BIC values, the exponential model $\rm E_{1}$ is 
chosen as the best statistical model for the  parameter values obtained from 
the MCMC analysis. The MCMC analysis yields the following results: the mean 
distance to the Galactic center (GC) is found to be 
 $R_{0}=8.034\pm0.012_{\rm stat}\pm0.586_{\rm sys}$ kpc; the bulge 
$\delta$ Scuti distribution has a triaxial shape with normalized ($a\equiv1$) 
axes ratios ($a:b:c$) as $1.000\pm 0.005:0.348\pm0.002:0.421\pm0.002$. 
Here $a$ is the semi-major axis lying in the Galactic plane and pointing 
towards us; $b$ and $c$ are the two semi-minor axes, the former lying in the 
Galactic plane and the later perpendicular to it.  Smaller values of $b$ as 
compared to $a$ obtained for Galacto-centric distances $R\ge 2.0$~kpc indicate the presence of a bar-like structure of the bulge with a  bar angle of $22^{\circ}.006\pm2^{\circ}.078$. 
\end{abstract}
\section{Introduction}
The central bulge of the Milky Way is distributed in a more or less spherical 
cloud of stars and has a triaxial structure which extends up to a few 
kilo-parsecs above and below the disk \citep{wegg13}.  It is the vertically 
thick, inner region of our Galaxy and contains a mixture of Population I and 
Population II stars \citep{free07}. The metallicities of the bulge stars vary  
significantly from quite metal-poor to very metal-rich: $-2\le [Fe/H]<0.5$~dex 
\citep{carr17}. It is the least understood and the  obscured part of our 
Galaxy as the stars are highly crowded in this region  which make it difficult 
to resolve individual stars. The presence of a lot of gas and dust along the 
line of sight causes the extinction to be very high which makes this part very 
difficult to study until recently with  the advent of technological 
advancements \citep{piet20}.

The $\delta$ Scuti stars are intermediate mass pulsating variable stars 
with spectral types between A3 and F2 located in the Cepheid instability strip 
on or above the main sequence  which correspond to surface temperatures in the range $\sim 8600~{\rm K}-6900~{\rm K}$ \citep{goup05,neme17,jays19,bedd20}. 
The definition of $\delta$ Scuti stars as defined in the literature includes 
both the populations: Population I and Population II \citep[and references 
therein]{neme17,sanc18,piet20,guzi21}.  Population I stars represent masses in 
the range $1.5~M_{\odot}< M<2.5~M_{\odot}$, whereas, Population II stars 
correspond to mass range $1~M_{\odot}<M<2~M_{\odot}$. SX Phe stars 
comprise Population II $\delta$ Scuti stars, however not all of them are 
found to be metal poor \citep{neme17,anto19}.  Furthermore, the distinction 
between the two populations is impossible to be made even with the highly 
precise Kepler data and hence the $\delta$ Scuti stars may be considered as stars having mixed populations \citep{neme17,guzi21}. They  act as `standard candles' for measuring distances within our Galaxy and to other 
nearby galaxies. Like RR Lyraes and Cepheids, they also obey a 
period-luminosity (PL) relation in specific pass bands 
\citep{mcna97,mcna11,ziaa19,poro21} which make them useful 
as `standard candles'. However, the scatter in the $V$-band PL relation is 
somewhat larger. Their absolute magnitudes are lower than those observed for 
the RR Lyraes and Cepheids \citep{mcna62,mcna97,mcna11}. They pulsate at high 
frequencies ($0.02~{\rm d}<P<0.25~{\rm d}$) with $V$-band amplitude in the 
range $0.003<A_{V}<{0.9}$~mag \citep{chan13,jays19}.

A large number of studies using direct and indirect methods in the recent past 
have been  carried out to determine the Galactic center (GC) distance $R_{0}$. 
All these studies lead to  a wide range of values as reported in the literature: $R_{0}=7.5-8.9$~kpc \citep{maja09,minn10,sait12,piet12,deka13,piet15,bhar17,cama18,maja18, reid19,griv19,griv20,griv21,boby21b}. One of the most recent and 
direct geometrical estimates of $R_{0}$ obtained by the GRAVITY team is 
$R_{0} [{\rm kpc}]=8.178\pm 0.013_{\rm stat}\pm0.022_{\rm sys}$. This estimate 
has been obtained by following the star S2 in its orbit around the massive 
black hole Sgr A$^{\star}$ for more than two decades both astrometrically and 
spectroscopically and does not rely on any calibration steps \citep{abut19}.    Accurate determination of $R_{0}$ plays an important role for reliable 
estimation of the parameters describing the structure of the Milky Way bulge 
as well as the overall structure of the Galaxy. Any  change in the value of 
$R_{0}$ has a widespread impact on Astronomy and Astrophysics 
\citep{reid93,blan16}.

A number of observational studies reveal that Milky Way has a asymmetric 
boxy-bulge \citep{weil94,binn97,skru06}. Evidence of a tilted bar in 
the inner Milky Way with its closer part at positive Galactic longitudes
 was found by \citet{blit91}. Mostly, the metal-rich stars like red clump (RC) 
stars trace a prominent tilted bar structure 
\citep{stan94,dwek95,stan97, babu05,ratt07,cao13,wegg13,simi17,simi21}. However,
metal-poor stars like RR Lyrae (RRL) stars show a remarkably
different spatial distribution compared to the RC stars \citep{gonz16}.
The RRLs do not trace a prominent Galactic bar but a more spheroidal 
centrally concentrated distribution with a slight elongation
in its center \citep{maja10,deka13,piet15,du20,griv20}. Similarly the
old metal-poor Type II Cepheids show a similar spheroidal structure as those 
of RRL stars \citep{deka19,griv21}. \citet{grad20} found that the older Mira 
variable stars do not show the evidence of a bar structure as the bar formed 
$8-9$~Gyr ago. But the intermediate-age Miras show a bar structure inclined at 
an angle $\sim 21^{\circ}$ with respect to the line joining the GC and the 
Sun. 

But there exist considerable differences involving the results of Galactic bar
structure as well as the value of inclination angle with respect to the 
line of sight \citep{ratt07,deka13,wegg13,cao13,piet12,piet15,simi17,deka19,du20,grad20,griv20,griv21,simi21}. The matter is still debated and a consensus has 
yet to be reached. At present, the axis ratio of the Galactic bar is 
constrained at $1:0.4:0.3$ \citep{gonz16}. Until recently, the inclination 
angle of the bar with respect to the line joining GC and Sun 
was constrained at a very large range $20^{\circ}-40^{\circ}$. On the other 
hand, smaller values of viewing angle for the bulge have also been reported by
\citet{griv20,griv21} for higher Galactic latitudes using RRLs and Type II 
Cepheids which are Population II stars, including \citet{pica04} and 
\citet{grad20}  using old Mira variables and infrared source counts of 
different stellar populations, respectively. The 
variations in the values of the inclination angle obtained in different 
studies may be attributed to the differences in the latitude ranges as 
well as different sources  of tracers used in the respective studies 
\citep{gonz16}. Hence further investigations are quite necessary to constrain 
the values of the geometrical and the viewing angle parameters of the Milky 
Way bulge.

The present study deals with the estimation of  $R_{0}$ and the geometrical 
parameters of the Milky Way bulge using the photometric data of a large 
number of  $\delta$ Scuti stars from the OGLE-IV database 
(the largest sample of $\delta$ Scuti stars known so far) based on ML and MCMC 
analyses. The study of  \citet{piet20}  reports $10111$ genuine $\delta$ 
Scuti-type pulsating stars detected in the OGLE (Optical Gravitational Lensing 
Experiment)-IV Galactic bulge fields covering $\sim 172$  square degrees. 
Most of the variables in the database ($9835$) are new discoveries. They are 
located at distances of several kpc from us and mainly belong to the 
intermediate-age and old populations of the bulge, since they are located at 
higher Galactic latitudes. This is consistent with the observations that most 
of the bulge stars are old \citep{rich13}. The objects that are detected by  
OGLE-IV are indeed $\delta$ Scuti stars are not only confirmed from the shapes 
of their light curves, period and peak-to-peak amplitudes but 
also from their locations in the colour-magnitude diagrams constructed for 
stars from the same field. The detected stars are found to reside in the area 
around the upper or middle Main-sequence in the observed diagrams 
\citep{piet20}. Besides these,  their distributions on the sky and 
color-magnitude diagrams  are found to be very similar to those of the bulge 
RRL variables which supplement the confirmation that these stars belong to 
the bulge \citep{piet15,piet20}.

The formulation  of  the ML analysis of the bulge parameters 
is based on the work of \citet{rast94} and \citet{griv19}. Recently released 
photometric $V$- and $I$-band data of $10092$ $\delta$ Scuti stars in the 
galactic bulge by the OGLE-IV team \citep{piet20} along with the updated PL 
relation of \citet{ziaa19} for these stars allows a unique opportunity to 
study the distance to the GC as well as the structural parameters of the bulge 
in great details. The empirical relation of \citet{ziaa19} may not provide 
accurate values of $M_{V}$ for an individual $\delta$ Scuti star but is 
suitable for the statistical analysis of a large number of the bulge $\delta$ 
Scuti stars discovered by OGLE-IV, which can serve as a powerful tracer to the 
underlying bulge structure.

\citet{rast94} for the first time used ML method to find $R_{0}$ using 
globular clusters in the halo of our Galaxy.  Based on the method 
developed by \citet{rast94}, recently, \citet{griv19,griv20,griv21} estimated 
the bulge parameters using globular clusters, RR Lyrae and  Type II Cepheid 
variables, respectively by approximating the number density of the the bulge
as a function of the Galacto-centric distance to obey a power-law 
distribution. Each of the parameters of the bulge in their 
studies were estimated using ML method keeping the other parameters fixed  
\citep{griv20}. Nonetheless, using likelihood slices is simple and useful for 
visualizing many parameter likelihood surfaces near its minimum but 
statistically misleading as they do not allow other parameters to vary 
\citep{bolk05}. In the present study, we use likelihood profile which allows 
to optimize the likelihood with respect to  all the model  parameters of the 
bulge. Furthermore, more refined values of these parameters and their 
statistical uncertainties are determined based on MCMC analysis. 
\begin{figure*}
\vspace{0.014\linewidth}
\begin{tabular}{ccc}
\vspace{+0.01\linewidth}
\resizebox{0.5\linewidth}{!}{\includegraphics*{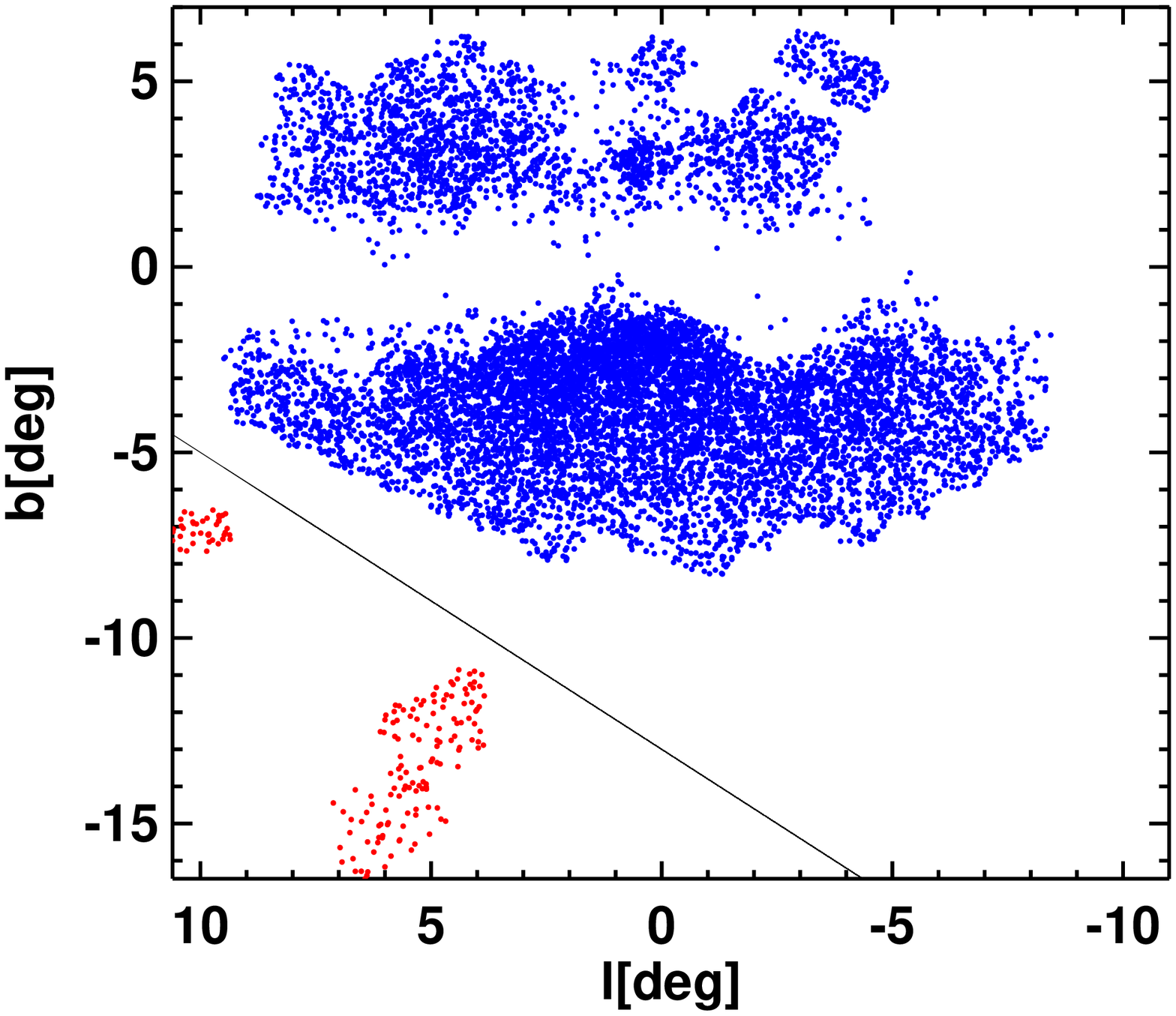}}&
\resizebox{0.5\linewidth}{!}{\includegraphics*{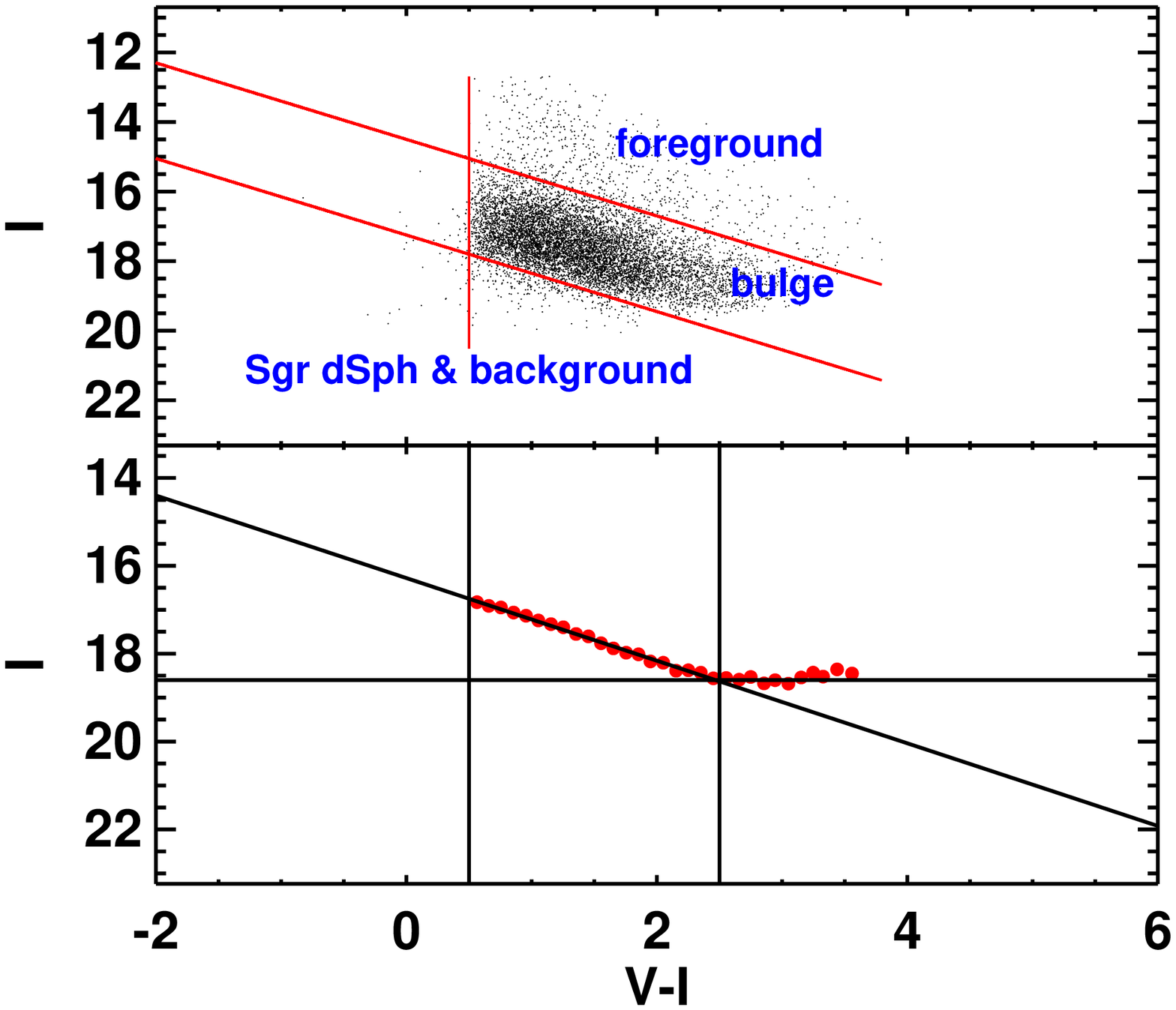}}\\
\vspace{-0.04\linewidth}
\end{tabular}
\caption{{\it Left panel}: Blue dots indicate the sample cleaned from the Sgr 
dSph galaxy and Galactic globular cluster M22 shown by red dots. 
{\it Upper right panel}: The selected sample for our analysis cleaned from 
foreground, background and other contaminant stars. {\it Lower right panel}: 
Binning cleaned data with bin size $0.1$~mag in the $V-I$ color show that the 
bulge sequence is linear down to $I\sim 18.6$ mag or $0.5\le V-I \le 2.5$~mag 
which is taken as the completeness limit for the present sample of bulge $\delta$ Scuti stars.}
\label{fig:Scuti}
\end{figure*}

The remaining paper is organized as follows: In Section~\ref{sec:data}, we 
discuss the selection and cleaning of the $\delta$ Scuti stars taken from 
the OGLE-IV database. Section~\ref{sec:structure} deals with the methodology 
and the results obtained in the present study. Finally, the summary and 
conclusion of the  present study are discussed in Section~\ref{sec:summary}. 
\section{Selection and Cleaning of Data}
\label{sec:data}
The photometric data of the $\delta$ Scuti stars are taken from the data
recently released by the  OGLE-IV survey obtained using the $1.3$ meter Warsaw 
telescope situated at Las Campanas Observatory in Chille \citep{piet20}. The 
OGLE-IV project has archived data of thousands of bulge $\delta$ Scuti stars 
from $2010$ to $2019$. The original OGLE sample consists of photometric data as 
well as the light curves  of  $10092$ $\delta$ Scuti stars in $V$ and 
$I$-bands. Seven fields around 
$(l,b)=(+5^{\circ},-13^\circ)$ and a single field at  
$(l,b)=(+10^{\circ},-7^\circ)$ cover the central part of the Sgr dSph galaxy
and Galactic globular cluster M22, respectively \citep{piet20}. The stars 
belonging to these regions are removed by drawing line: $b=0.8l-13.0$ as shown 
in left panel of 
Fig.~\ref{fig:Scuti}. Then a clean sample of data free from the possible 
contamination of background and foreground sources is selected by closely 
following the selection criterion as prescribed by \citet{piet15}.
The foreground (Halo and Disk stars; cf. \citet{piet20}) 
and the background stars are removed by drawing lines at $I < 14.5 + 1.1(V-I)$,
$I>17.25 + 1.1(V-I)$ and $V-I <0.5$ as shown in the upper right panel of 
Fig.~\ref{fig:Scuti}. This step reduces the number of $\delta$ Scuti stars for 
further analysis to $8,356$. For finding the photometric completeness of 
the sample of $\delta$ Scuti stars, we resort to the procedure as given by 
\citet{piet15}. The cleaned data are binned with a bin size $0.1$ as shown in 
the lower right panel of Fig.~\ref{fig:Scuti}. It can be seen from the figure 
that the mean brightness of the bulge section is linear for 
$I\le18.6$ or $0.5\le V-I\le2.5$. Hence following \citet{piet15}, we assume 
that our sample of $\delta$ Scuti stars is complete down to $I=18.6$~mag. Hence 
stars having $I\ge 18.6$~mag are removed from the clean sample which reduces 
the final number of $\delta$ Scuti stars to $7,440$ for further analysis.
\section{Methodology}
\label{sec:structure}
\subsection{Distance to individual star}
$\delta$ Scuti stars obey the following PL relation \citep{ziaa19}:
\begin{align}
\label{eq:abs}
M_{V}=&(-2.94\pm0.06)\log{P}-(1.34\pm0.06),~~~~\sigma_{\rm in}=0.34,
\end{align}
where $M_{V}$ is the absolute magnitude in $V$-band; $P$, the period of the 
star in days and $\sigma_{\rm in}=0.34$~mag denotes the intrinsic 
dispersion of the PL relation. The intrinsic dispersion is not mentioned in 
\citet{ziaa19} and has been obtained through a private communication. This 
empirical relation is based on more than $1100$ $\delta$ Scuti stars with good 
Gaia DR2 parallaxes in the Kepler field. 

The absolute magnitudes are also calculated using the PL relation of \citet{mcna11}:
\begin{align}
\label{eq:abs1}
M_{V}=&(-2.89\pm0.13)\log{P}-(1.31\pm0.10),~~~~\sigma_{\rm in}=0.07,
\end{align}

\begin{figure*}
\vspace{0.014\linewidth}
\begin{tabular}{ccc}
\vspace{+0.01\linewidth}
  \resizebox{0.5\linewidth}{!}{\includegraphics*{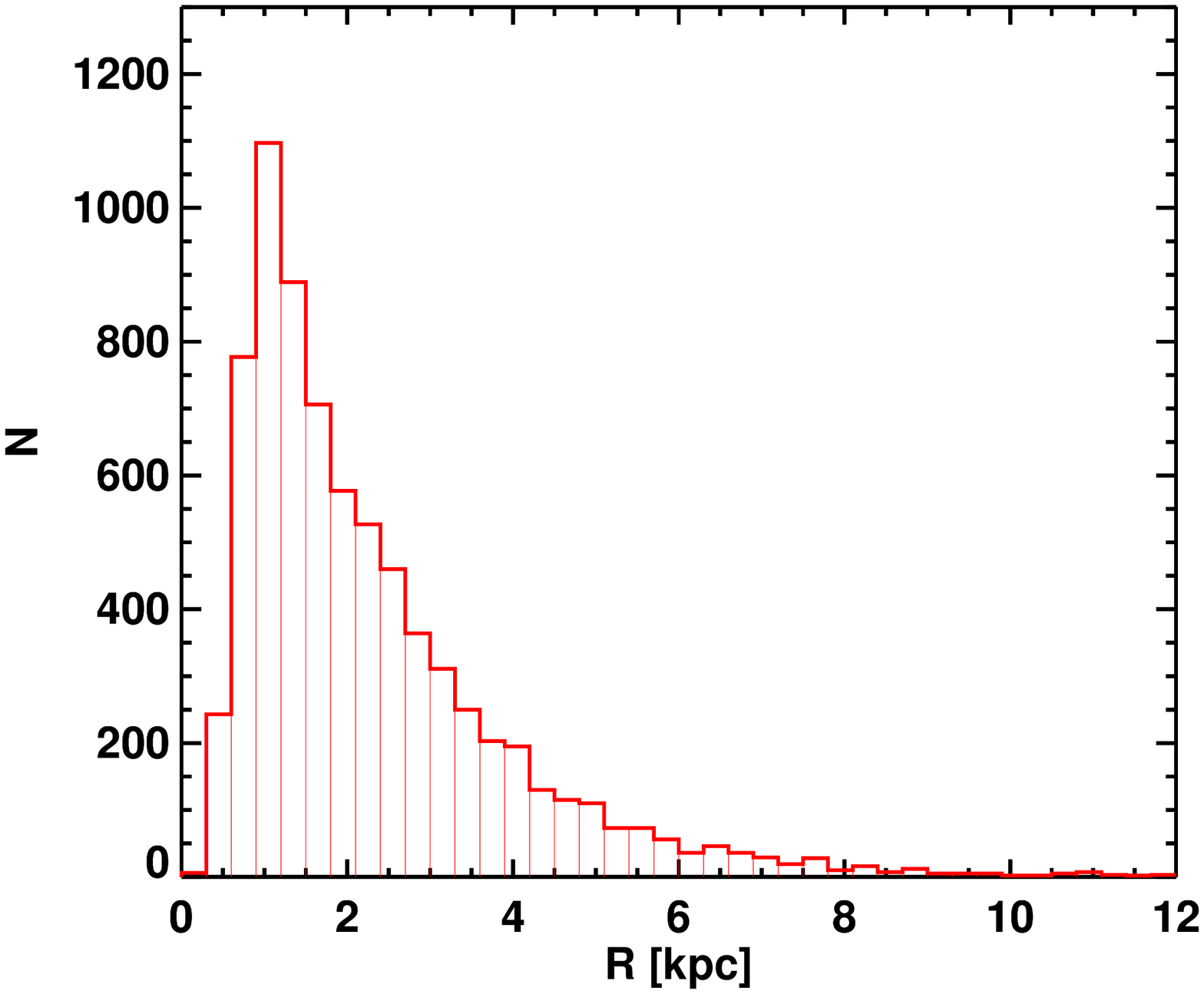}}&
  \resizebox{0.5\linewidth}{!}{\includegraphics*{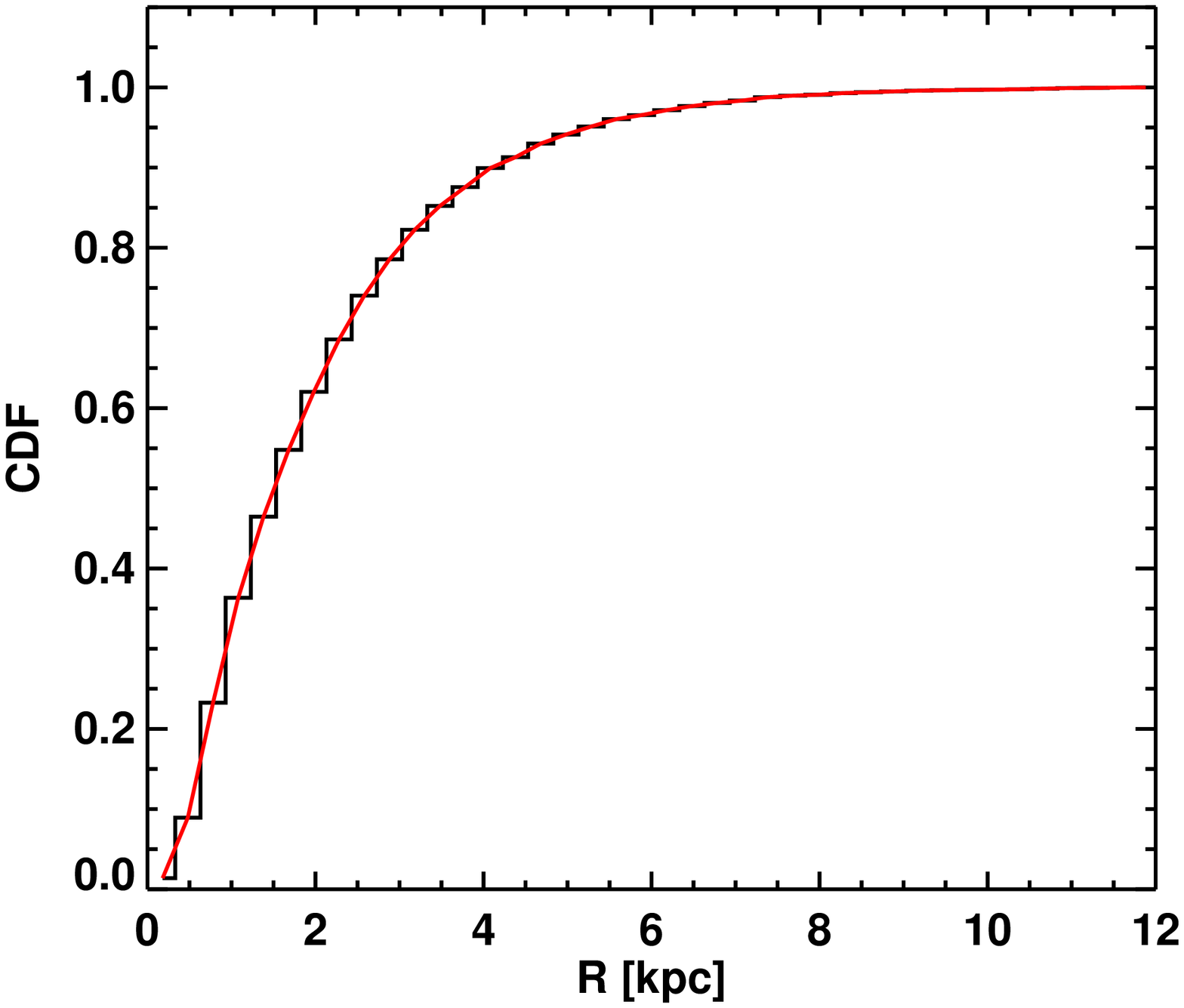}}\\
\vspace{-0.04\linewidth}
\end{tabular}
\caption{{\it Left panel}: The distribution of Galacto-centric distances of 
$\delta$ Scuti stars. {\it Right panel}: Cumulative distribution of 
Galacto-centric distances of $\delta$ Scuti stars. }
\label{fig:gal_dist}
\end{figure*}

The true distance modulus is obtained using 
\begin{align}
\label{eq:dis_ex}
\mu_{0}=\mu_{V}-R_{V}E(B-V),
\end{align}
where $\mu_{0}$ is the true distance modulus corrected for extinction; 
$\mu_{V}=\overline{m}_{V}-M_{V}$, the apparent distance modulus 
in the $V$-band. The values of $P$ and $\overline{m}_{V}$ are taken from the 
OGLE-IV database; $R_{V}$, the ratio of total to selective absorption in the
V-band and $E(B-V)$ is the interstellar reddening along the line of sight.
$R_{V}$ is taken to be $2.5$ \citep{nata13}. $E(B-V)$ is 
estimated as follows:
We know that
\begin{align}
A_{V}=R_{V}E(B-V)
\label{eq:av}
\end{align}
and from \citet{card89} extinction law \citep{gonz12_jks}:
\begin{align}
A_{J}=1.692E(J-K_{s}).
\label{eq:aj}
\end{align}
Dividing equation~(\ref{eq:aj}) by equation~(\ref{eq:av}),
\begin{align}
\frac{A_{J}}{A_{V}}=&\frac{1.692E(J-K_{s})}{R_{V}E(B-V)} \\
\Rightarrow E(B-V)=&\frac{1.692E(J-K_{s})}{R_{V}\left(\frac{A_{J}}{A_{V}}\right)}.
\label{eq:avj}
\end{align}
Here $\frac{A_{J}}{A_{V}}=0.26$, as obtained using \citet{card89} extinction 
law and $E(J-K_{S})$ is obtained from the \citet{gonz12} reddening map. 
Finally the distance is calculated using the following relation:
\begin{align}
d=&10^{0.2\mu_{0}+1}.
\end{align}
\begin{figure}
\includegraphics[width=0.5\textwidth,keepaspectratio]{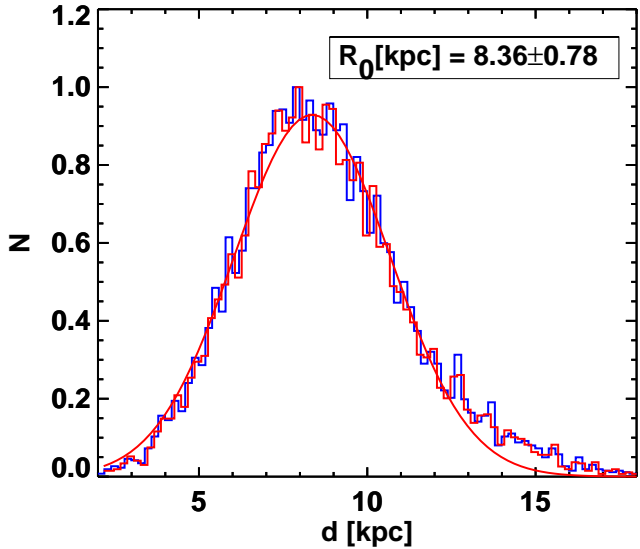}
\caption{Histogram plot of the distance distribution of $\delta$ Scuti stars
without and with cone-effect corrections as represented by blue and red colours, respectively. Solid line over plotted in red color denotes the Gaussian fit to 
the distance distribution corrected from the cone-effect.}
\label{fig:helio}
\end{figure}
\begin{figure*}
\vspace{0.014\linewidth}
\begin{tabular}{ccc}
\vspace{+0.01\linewidth}
  \resizebox{0.32\linewidth}{!}{\includegraphics*{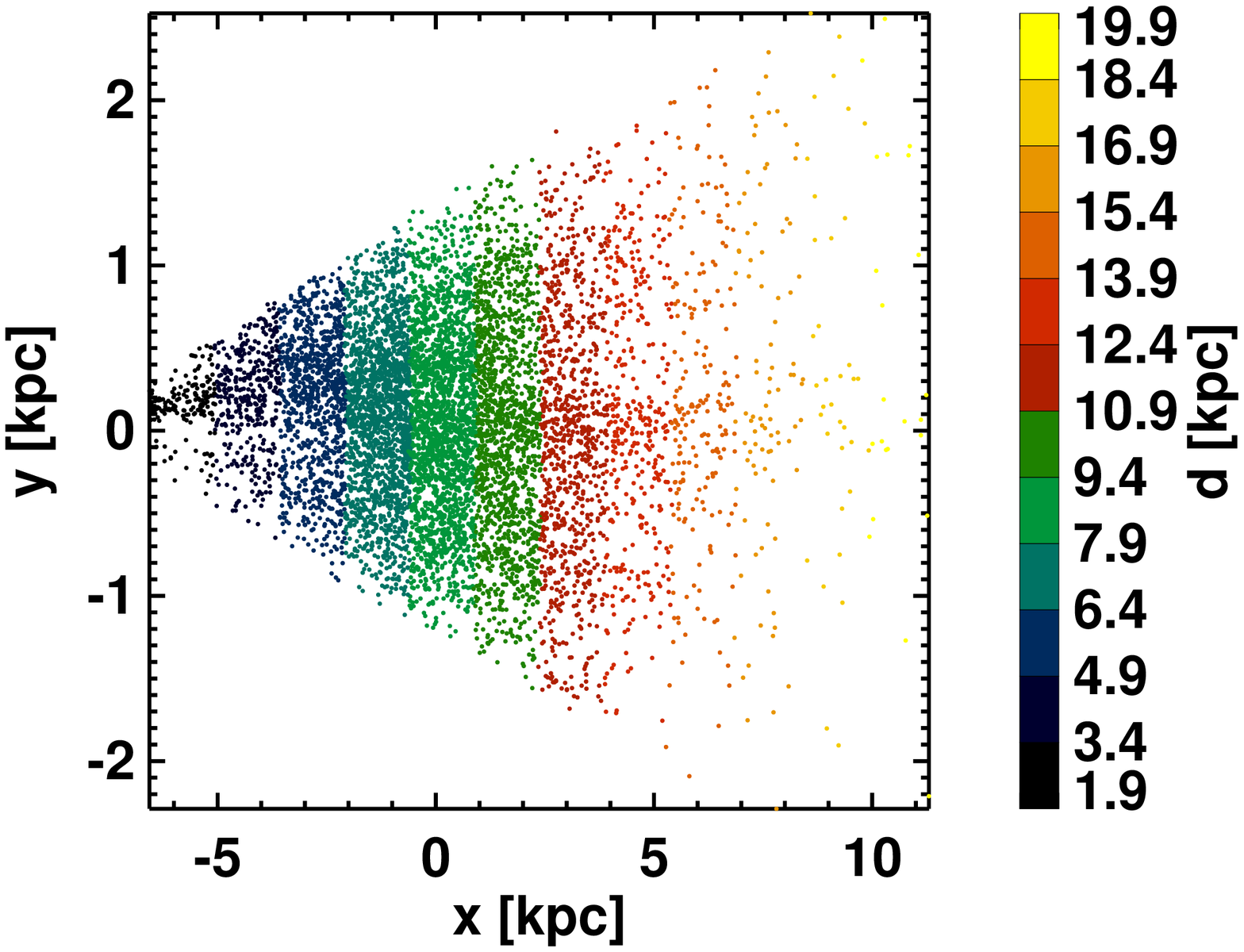}}&
\resizebox{0.32\linewidth}{!}{\includegraphics*{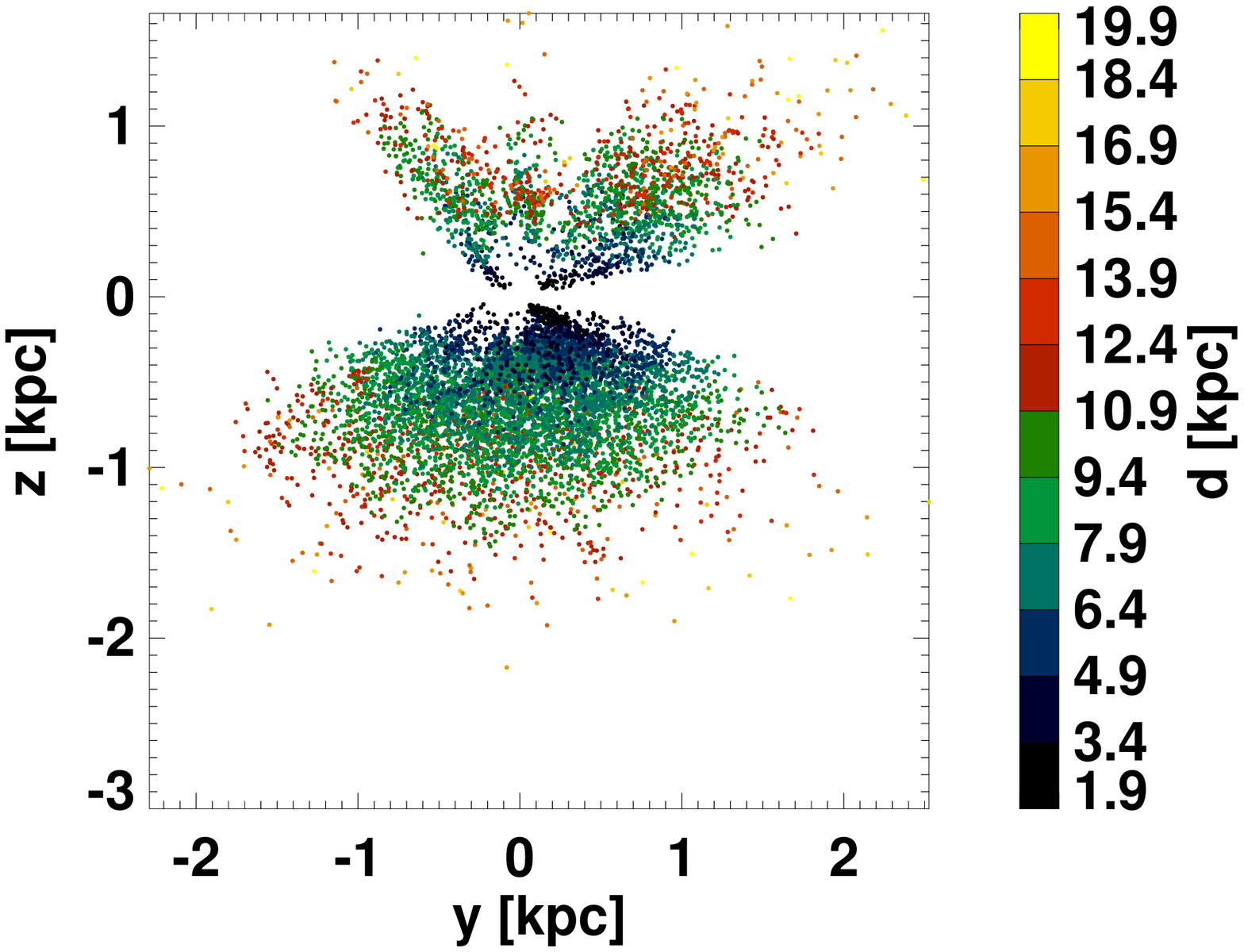}}&
  \resizebox{0.32\linewidth}{!}{\includegraphics*{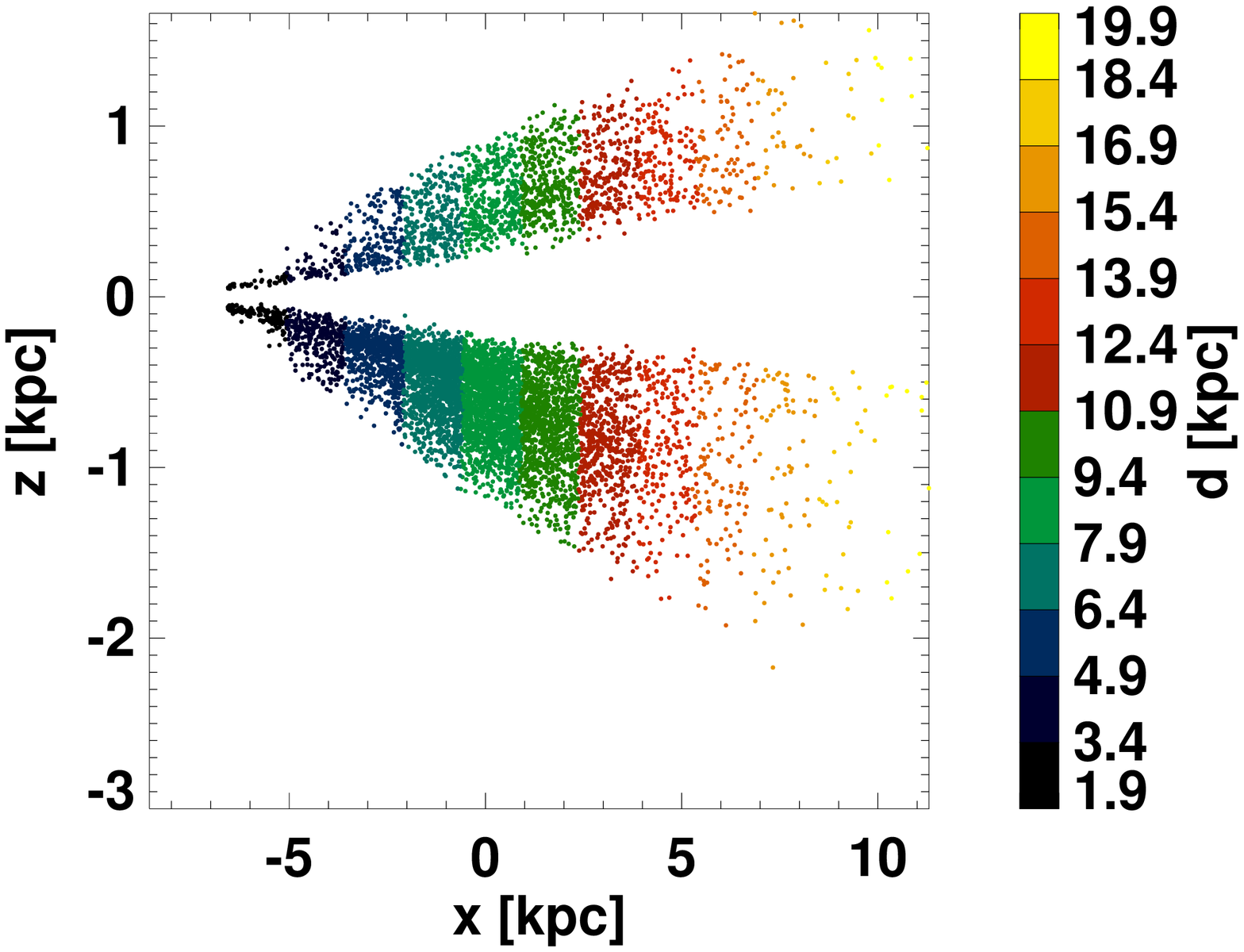}}\\
\vspace{-0.04\linewidth}
\end{tabular}
\caption{ The projected distribution of bulge $\delta$ Scuti stars in Cartesian
coordinate with the colour bar of distances.}
\label{fig:xyz_plot}
\end{figure*}
A histogram plot of the distance is shown in Fig.~\ref{fig:helio}. 
The distribution can be well represented by a Gaussian fit. 
The distribution is first  corrected from the 
cone-effect and then fitted with a Gaussian profile. The peak of the 
distribution is found to be at $8.36\pm 0.78$~kpc which is taken as a 
preliminary estimate of $R_{0}$ for obtaining the radial number density 
profile of stars from the GC. Since the distance to individual stars is known, 
the following 
coordinate transformations are then used to convert Galactic coordinates into 
the Galacto-centric Cartesian coordinates: 
\begin{align*}
x=& d\cos{b}\cos{l}-R_{0} \\
y=&d\cos{b}\sin{l} \\
z= &d\sin{b}, 
\end{align*}
where $x$-axis is directed along the Galactic anticenter; $y$-axis directed 
along the direction of Galactic rotation and  $z$-axis is directed towards 
Galactic north. Colour-bar plots of the distribution of stars in the $xy, yz$
and $xz$-planes are shown in Fig.~\ref{fig:xyz_plot}. The Galacto-centric 
distance is then calculated using
\begin{align}
R=&\sqrt{x^{2}+y^{2}+(z+z_{0})^2}.
\end{align}
Here $z_{0}$ is the Sun's distance from the Galactic plane and is taken as  
$z_{0}=(20\pm2)$~pc \citep{griv21}. Number density distribution of $R$ of the
selected sample of $\delta$ Scuti stars  and the corresponding cumulative 
distribution function (CDF) are shown in the left and right panels of 
Fig.~\ref{fig:gal_dist}, respectively. 
\subsection{Maximum Likelihood Parameter Estimation}
We approximate the bulge as an ellipsoid given by 
\begin{align}
\frac{x^{2}}{a^{2}}+\frac{y^{2}}{b^{2}}+\frac{z^{2}}{c^{2}}=&1,
\end{align}
where $a,b,c$ are the normalized axes ($a\equiv1$) of the ellipsoid with 
$a$, the major axis lying in the Galactic plane and  pointing towards us and
the two minor axes $b$ and $c$ lying in the Galactic plane and perpendicular 
to the Galactic plane, respectively. We assume the general case 
($a \neq b \neq c$). The coordinate system $(x,y,z)$ is rotated in 
clockwise direction through $xy$-plane by an angle $\theta$ as 
\begin{align}
\begin{pmatrix}
X\\
Y\\
Z
\end{pmatrix} =&
\begin{pmatrix}
\cos{\theta} & -\sin{\theta} & 0\\
\sin{\theta} & \cos{\theta} & 0\\
	0 & 0 & 1
\end{pmatrix}
\begin{pmatrix}
x\\
y\\
z
\end{pmatrix}.
\end{align}
\begin{table*}
\caption{Geometrical parameters of the bulge for three different density 
distribution laws using the PL relation of \citet{ziaa19} with the given 
$\sigma_{\rm in}=0.34$ mag.}
\label{table:tab34}
\begin{center}
\begin{tabular}{l|c|c|c|c|c|c|r}  
\hline 
Model & $R_{\rm min}-R_{\rm max}$~[kpc] & $R_{0}$~[kpc] & $\theta~[^{\circ}]$ & $a:b:c$ \\ \hline 
${\rm E_{1}}$ & $0.05-1.5$ & $8.310\pm0.037$ & $13.681\pm5.621$ & $1.000\pm0.016:0.702\pm0.011:0.877\pm0.014$ \\ 
$~$ & $0.05-2.0 $ & $8.024\pm0.037 $ & $17.523\pm5.327 $ & $1.000\pm0.014:0.524\pm0.007:0.640\pm0.009$ \\ 
$~$ & $0.05-2.5$ & $8.033\pm0.029$ & $21.156\pm4.864$ & $1.000\pm0.013:0.425\pm0.005:0.515\pm0.006$ \\ 
$~$ & $0.05-3.0$ & $8.013\pm0.027$ & $22.330\pm4.267$ & $1.000\pm0.012:0.360\pm0.004:0.435\pm0.005$ \\ 
$~$ & $0.05-3.5$ & $8.044\pm0.025$ & $23.603\pm4.507$ & $1.000\pm0.011:0.323\pm0.004:0.390\pm0.004$ \\ 
$~$ & $0.05-4.0$ & $8.050\pm0.025$ & $23.954\pm4.471$ & $1.000\pm0.011:0.296\pm0.003:0.359\pm0.004$ \\ \hline 
${\rm E_{2}}$ & $0.05-1.5 $ & $8.047\pm0.032$ & $17.794\pm5.731$ & $1.000\pm0.016:0.674\pm0.011:0.741\pm0.011$ \\ 
$~$ & $0.05-2.0 $ & $8.013\pm0.027 $ & $20.718\pm4.348$ & $1.000\pm0.014:0.508\pm0.007:0.546\pm0.007$ \\ 
$~$ & $0.05-2.5$ & $8.001\pm0.025$ & $23.019\pm3.796$ & $1.000\pm0.012:0.412\pm0.005:0.440\pm0.005$ \\ 
$~$ & $0.05-3.0$ & $7.958\pm0.023$ & $22.443\pm3.173$ & $1.000\pm0.011:0.350\pm0.004:0.372\pm0.004$ \\ 
$~$ & $0.05-3.5$ & $7.973\pm0.022$ & $22.872\pm3.117$ & $1.000\pm0.011:0.314\pm0.003:0.335\pm0.003$ \\
$~$ & $0.05-4.0$ & $7.973\pm0.021$ & $22.745\pm3.045$ & $1.000\pm0.011:0.289\pm0.003:0.309\pm0.003$ \\ \hline
$\rm G$ & $0.05-1.5$ & $8.212\pm0.019$ & $14.542\pm6.546$ & $1.000\pm0.013:0.705\pm0.009:0.712\pm0.008$ \\ 
$~$ & $0.05-2.0$ & $8.208\pm0.017$ & $20.285\pm6.183$ & $1.000\pm0.011:0.531\pm0.006:0.526\pm0.005$ \\ 
$~$ & $0.05-2.5$ & $8.221\pm0.015$ & $27.389\pm6.089$ & $1.000\pm0.010:0.431\pm0.004:0.424\pm0.004$ \\ 
$~$ & $0.05-3.0$ & $8.208\pm0.014$ & $27.650\pm5.242$ & $1.000\pm0.009:0.365\pm0.003:0.359\pm0.003$ \\ 
$~$ & $0.05-3.5$ & $8.231\pm0.013$ & $30.764\pm5.944$ & $1.000\pm0.009:0.327\pm0.003:0.323\pm0.003$ \\  
$~$ & $0.05-4.0$ & $8.239\pm0.010$ & $31.491\pm0.562$ & $1.000\pm0.006:0.300\pm0.006:0.297\pm0.006$ \\ \hline
\end{tabular}
\end{center}
\end{table*}

\begin{table*}
\caption{Geometrical parameters of the bulge for three different density 
distribution laws using the PL relation of \citet{ziaa19} assuming
$\sigma_{\rm in}=0.2$ mag.}
\label{table:tab2}
\begin{center}
\begin{tabular}{l|c|c|c|c|c|c|r}  
\hline 
Model & $R_{\rm min}-R_{\rm max}$~[kpc] & $R_{0}$~[kpc] & $\theta~[^{\circ}]$ & $a:b:c$ \\ \hline 
${\rm E_{1}}$ & $0.05-1.5$ & $8.082\pm0.036$ & $11.183\pm5.373$ & $1.000\pm0.016:0.717\pm0.011:0.897\pm0.014$ \\ 
$~$ & $0.05-2.0 $ & $8.118\pm0.031 $ & $17.808\pm5.163 $ & $1.000\pm0.013:0.537\pm0.007:0.657\pm0.009$ \\ 
$~$ & $0.05-2.5$ & $8.187\pm0.028$ & $27.285\pm5.887$ & $1.000\pm0.012:0.442\pm0.005:0.539\pm0.006$ \\ 
$~$ & $0.05-3.0$ & $8.175\pm0.026$ & $29.173\pm5.276$ & $1.000\pm0.011:0.382\pm0.004:0.465\pm0.005$ \\ 
$~$ & $0.05-3.5$ & $8.178\pm0.025$ & $29.756\pm5.113$ & $1.000\pm0.011:0.344\pm0.004:0.419\pm0.005$ \\ 
$~$ & $0.05-4.0$ & $8.181\pm0.024$ & $30.395\pm5.077$ & $1.000\pm0.011:0.322\pm0.003:0.391\pm0.004$ \\ \hline 
$\rm E_{2}$ & $0.05-1.5$ & $8.079\pm0.031$ & $13.552\pm4.599$ & $1.000\pm0.015:0.686\pm0.010:0.757\pm0.011$ \\ 
$~$ & $0.05-2.0$ & $8.097\pm0.026$ & $19.283\pm4.174$ & $1.000\pm0.013:0.519\pm0.007:0.559\pm0.007$ \\ 
$~$ & $0.05-2.5$ & $8.138\pm0.024$ & $26.021\pm4.303$ & $1.000\pm0.012:0.429\pm0.005:0.460\pm0.005$ \\ 
$~$ & $0.05-3.0$ & $8.118\pm0.023$ & $26.865\pm3.807$ & $1.000\pm0.011:0.370\pm0.004:0.396\pm0.004$ \\ 
$~$ & $0.05-3.5$ & $8.114\pm0.022$ & $26.993\pm3.549$ & $1.000\pm0.011:0.334\pm0.004:0.358\pm0.004$ \\  
$~$ & $0.05-4.0$ & $8.110\pm0.021$ & $26.893\pm3.459$ & $1.000\pm0.011:0.311\pm0.003:0.335\pm0.003$ \\ \hline
${\rm G}$ & $0.05-1.5 $ & $8.277\pm0.018$ & $9.273\pm5.503$ & $1.000\pm0.012:0.719\pm0.008:0.734\pm0.008$ \\ 
$~$ & $0.05-2.0 $ & $8.305\pm0.016 $ & $17.344\pm6.136$ & $1.000\pm0.010:0.542\pm0.005:0.539\pm0.005$ \\ 
$~$ & $0.05-2.5$ & $8.342\pm0.014$ & $32.884\pm7.593$ & $1.000\pm0.009:0.446\pm0.004:0.443\pm0.004$ \\ 
$~$ & $0.05-3.0$ & $8.333\pm0.013$ & $35.086\pm6.690$ & $1.000\pm0.009:0.383\pm0.003:0.380\pm0.003$ \\ 
$~$ & $0.05-3.5$ & $8.338\pm0.013$ & $36.448\pm6.715$ & $1.000\pm0.008:0.344\pm0.003:0.342\pm0.003$ \\
$~$ & $0.05-4.0$ & $8.340\pm0.013$ & $37.162\pm6.781$ & $1.000\pm0.008:0.319\pm0.003:0.318\pm0.002$ \\ \hline
\end{tabular}
\end{center}
\end{table*}

\begin{table*}
\caption{Geometrical parameters of the bulge for three different density 
distribution laws using the PL relation of \citet{ziaa19} assuming 
$\sigma_{\rm in}=0.1$ mag.}
\label{table:tab1}
\begin{center}
\begin{tabular}{l|c|c|c|c|c|c|r}  
\hline 
Model & $R_{\rm min}-R_{\rm max}$~[kpc] & $R_{0}$~[kpc] & $\theta~[^{\circ}]$ & $a:b:c$ \\ \hline 
${\rm E_{1}}$ & $0.05-1.5$ & $8.167\pm0.035$ & $12.852\pm6.271$ & $1.000\pm0.015:0.712\pm0.011:0.900\pm0.013$ \\ 
$~$ & $0.05-2.0 $ & $8.193\pm0.030 $ & $22.322\pm6.085 $ & $1.000\pm0.013:0.548\pm0.007:0.672\pm0.008$ \\ 
$~$ & $0.05-2.5$ & $8.194\pm0.027$ & $25.495\pm5.546$ & $1.000\pm0.012:0.450\pm0.005:0.553\pm0.007$ \\ 
$~$ & $0.05-3.0$ & $8.218\pm0.026$ & $27.024\pm5.735$ & $1.000\pm0.011:0.398\pm0.004:0.485\pm0.005$ \\ 
$~$ & $0.05-3.5$ & $8.224\pm0.025$ & $29.395\pm5.673$ & $1.000\pm0.011:0.360\pm0.004:0.438\pm0.005$ \\ 
$~$ & $0.05-4.0$ & $8.216\pm0.024$ & $30.508\pm5.323$ & $1.000\pm0.011:0.333\pm0.004:0.405\pm0.004$ \\ \hline 
$\rm E_{2}$ & $0.05-1.5$ & $8.146\pm0.030$ & $16.277\pm5.084$ & $1.000\pm0.015:0.686\pm0.010:0.765\pm0.011$ \\ 
$~$ & $0.05-2.0$ & $8.150\pm0.026$ & $23.104\pm4.429$ & $1.000\pm0.013:0.529\pm0.007:0.572\pm0.007$ \\ 
$~$ & $0.05-2.5$ & $8.138\pm0.024$ & $24.560\pm3.947$ & $1.000\pm0.012:0.437\pm0.005:0.473\pm0.005$ \\ 
$~$ & $0.05-3.0$ & $8.147\pm0.022$ & $24.787\pm3.847$ & $1.000\pm0.011:0.387\pm0.004:0.417\pm0.004$ \\ 
$~$ & $0.05-3.5$ & $8.146\pm0.021$ & $25.936\pm3.694$ & $1.000\pm0.011:0.350\pm0.004:0.376\pm0.004$ \\  
$~$ & $0.05-4.0$ & $8.136\pm0.021$ & $26.445\pm3.471$ & $1.000\pm0.011:0.323\pm0.003:0.348\pm0.003$ \\ \hline
${\rm G}$ & $0.05-1.5 $ & $8.313\pm0.018$ & $9.870\pm5.861$ & $1.000\pm0.012:0.718\pm0.008:0.741\pm0.008$ \\ 
$~$ & $0.05-2.0 $ & $8.336\pm0.015 $ & $21.474\pm6.528$ & $1.000\pm0.010:0.552\pm0.005:0.551\pm0.005$ \\ 
$~$ & $0.05-2.5$ & $8.341\pm0.014$ & $26.522\pm6.283$ & $1.000\pm0.009:0.453\pm0.004:0.454\pm0.004$ \\ 
$~$ & $0.05-3.0$ & $8.358\pm0.013$ & $28.474\pm6.903$ & $1.000\pm0.009:0.398\pm0.003:0.396\pm0.003$ \\ 
$~$ & $0.05-3.5$ & $8.364\pm0.013$ & $33.453\pm6.929$ & $1.000\pm0.008:0.358\pm0.003:0.354\pm0.003$ \\
$~$ & $0.05-4.0$ & $8.360\pm0.013$ & $35.108\pm6.719$ & $1.000\pm0.008:0.327\pm0.003:0.326\pm0.002$ \\ \hline
\end{tabular}
\end{center}
\end{table*}

\begin{table*}
\caption{Geometrical parameters of the bulge for three different density 
distribution laws using the PL relation of \citet{mcna11} with the given
$\sigma_{\rm in}=0.07$ mag.}
\label{table:tab07}
\begin{center}
\begin{tabular}{l|c|c|c|c|c|c|r}  
\hline 
Model & $R_{\rm min}-R_{\rm max}$~[kpc] & $R_{0}$~[kpc] & $\theta~[^{\circ}]$ & $a:b:c$ \\ \hline 
${\rm E_{1}}$ & $0.05-1.5$ & $8.327\pm0.035$ & $14.568\pm7.788$ & $1.000\pm0.015:0.731\pm0.011:0.925\pm0.014$ \\ 
$~$ & $0.05-2.0 $ & $8.313\pm0.030 $ & $24.334\pm6.479 $ & $1.000\pm0.013:0.551\pm0.007:0.681\pm0.009$ \\ 
$~$ & $0.05-2.5$ & $8.313\pm0.027$ & $27.997\pm6.011$ & $1.000\pm0.012:0.464\pm0.005:0.567\pm0.007$ \\ 
$~$ & $0.05-3.0$ & $8.348\pm0.025$ & $31.876\pm6.431$ & $1.000\pm0.011:0.405\pm0.005:0.494\pm0.006$ \\ 
$~$ & $0.05-3.5$ & $8.342\pm0.024$ & $31.012\pm6.059$ & $1.000\pm0.011:0.365\pm0.004:0.445\pm0.005$ \\ 
$~$ & $0.05-4.0$ & $8.341\pm0.024$ & $31.506\pm5.918$ & $1.000\pm0.011:0.339\pm0.004:0.413\pm0.004$ \\ \hline 
${\rm E_{2}}$ & $0.05-1.5 $ & $8.319\pm0.030$ & $18.104\pm6.523$ & $1.000\pm0.015:0.701\pm0.010:0.783\pm0.011$ \\ 
$~$ & $0.05-2.0 $ & $8.271\pm0.026 $ & $24.730\pm4.776$ & $1.000\pm0.013:0.530\pm0.007:0.578\pm0.007$ \\ 
$~$ & $0.05-2.5$ & $8.261\pm0.024$ & $26.672\pm4.266$ & $1.000\pm0.012:0.449\pm0.005:0.484\pm0.005$ \\ 
$~$ & $0.05-3.0$ & $8.278\pm0.022$ & $28.257\pm4.263$ & $1.000\pm0.011:0.392\pm0.004:0.423\pm0.004$ \\ 
$~$ & $0.05-3.5$ & $8.261\pm0.021$ & $26.758\pm3.880$ & $1.000\pm0.011:0.353\pm0.004:0.381\pm0.004$ \\
$~$ & $0.05-4.0$ & $8.257\pm0.021$ & $26.643\pm3.687$ & $1.000\pm0.011:0.328\pm0.003:0.354\pm0.004$ \\ \hline
$\rm G$ & $0.05-1.5$ & $8.452\pm0.018$ & $10.747\pm7.314$ & $1.000\pm0.012:0.733\pm0.008:0.756\pm0.008$ \\ 
$~$ & $0.05-2.0$ & $8.440\pm0.015$ & $23.152\pm6.629$ & $1.000\pm0.010:0.554\pm0.005:0.557\pm0.005$ \\ 
$~$ & $0.05-2.5$ & $8.446\pm0.014$ & $29.501\pm6.628$ & $1.000\pm0.009:0.466\pm0.004:0.463\pm0.004$ \\ 
$~$ & $0.05-3.0$ & $8.470\pm0.013$ & $35.505\pm7.775$ & $1.000\pm0.009:0.404\pm0.003:0.402\pm0.003$ \\ 
$~$ & $0.05-3.5$ & $8.467\pm0.013$ & $33.201\pm7.226$ & $1.000\pm0.008:0.361\pm0.003:0.360\pm0.003$ \\  
$~$ & $0.05-4.0$ & $8.470\pm0.013$ & $33.868\pm7.296$ & $1.000\pm0.008:0.333\pm0.003:0.332\pm0.003$ \\ \hline
\end{tabular}
\end{center}
\end{table*}
\begin{figure*}
\begin{center}
\includegraphics[width=1\textwidth,keepaspectratio]{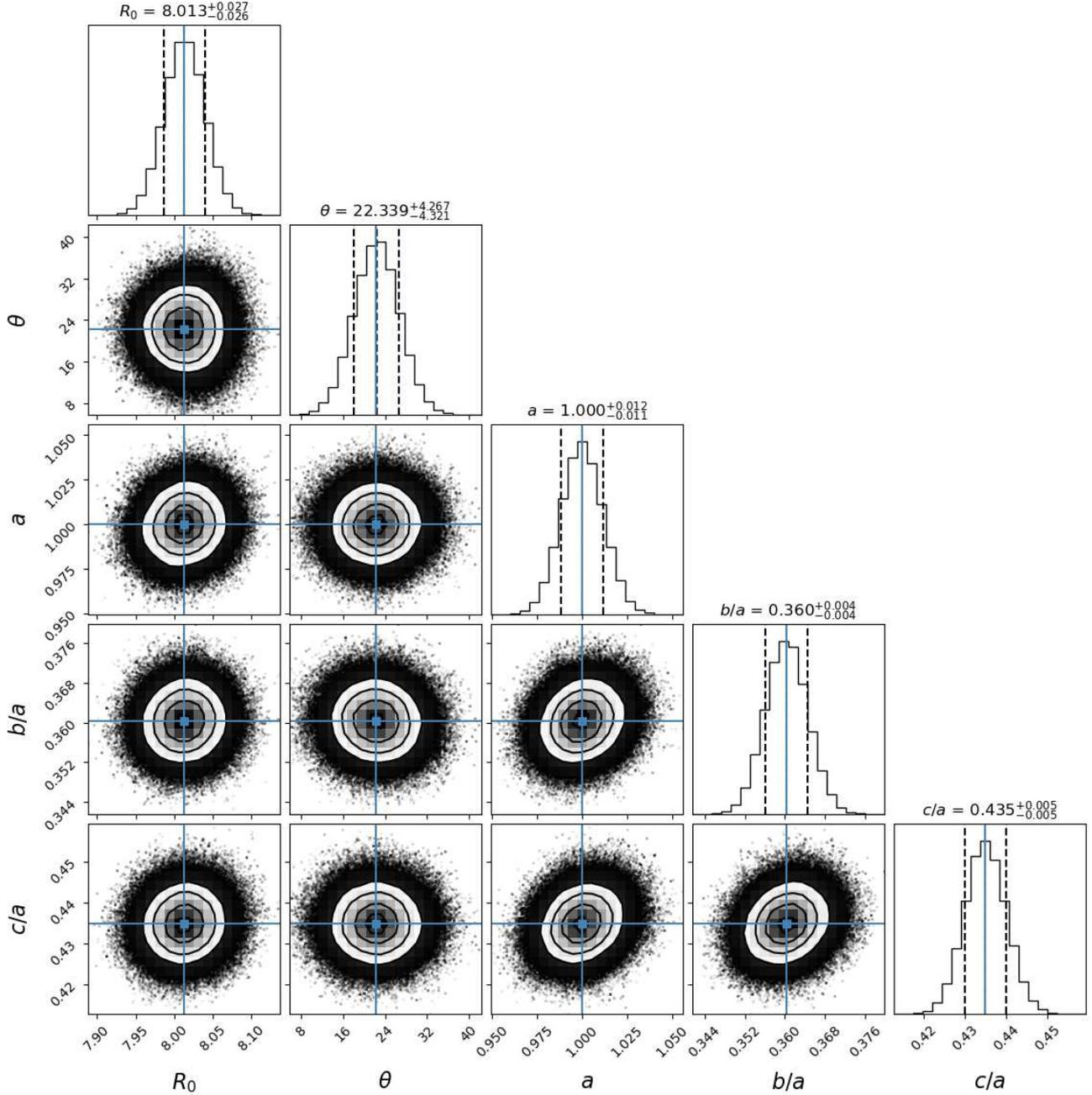}
\caption{Marginalized posterior distribution and uncertainties of the model parameters obtained from the MCMC analysis for the $\rm E_{1}$-model with 
$R_{\rm max}=3$~kpc considering \citet{ziaa19} PL relation with 
$\sigma_{\rm in}=0.34$ mag. The statistical uncertainties listed here are 
measured with the sample mean and $16$th (down) and $84$th(up) percentile 
differences. The black dashed vertical lines on each histogram represent the 
$16$th, $50$th and $84$th percentiles, respectively. This plot uses the corner 
package \citep{corn16}.}
\label{fig:emc}
\end{center}
\end{figure*}

\begin{table*}
\caption{BIC and AIC  values for the optimized parameters of  different models.}
\label{table:tab_bc}
\begin{center}
\begin{tabular}{l|c|c|c|c|c|c|r} 
\hline 
PL relation & $\sigma_{\rm in}$ & Model & $R_{\rm min}-R_{\rm max}$~[kpc] & BIC & AIC \\ \hline 
\citet{ziaa19}&$0.34$&$\rm E_{1}$ & $0.05-3.0 $ & $27335.241$ & $27345.241$ \\ 
& &$\rm E_{2}$ & $0.05-3.0$ & $34735.598$ & $34702.382$ \\ 
& &$\rm G$ & $0.05-3.0$ & $40794.816$ & $40761.599$ \\ \hline
\citet{ziaa19}&$0.2~(\text{assumed})$&$\rm E_{1}$ & $0.05-3.0 $ & $27494.606$ & $27504.606$ \\ 
& &$\rm E_{2}$ & $0.05-3.0$ & $36617.056$ & $36583.592$ \\ 
& &$\rm G$ & $0.05-3.0$ & $41609.247$ & $41575.782$ \\ \hline
\citet{ziaa19}&$0.1~(\text{assumed})$&$\rm E_{1}$ & $0.05-3.0 $ & $26937.706$ & $26947.706$ \\ 
& &$\rm E_{2}$ & $0.05-3.0$ & $36992.276$ & $36958.719$ \\ 
& &$\rm G$ & $0.05-3.0$ & $41126.608$ & $41093.050$ \\ \hline
\citet{mcna11}&$0.07$&$\rm E_{1}$ & $0.05-3.0 $ & $27847.405$ & $27857.405$ \\ 
& &$\rm E_{2}$ & $0.05-3.0$ & $37441.682$ & $37408.122$ \\ 
& &$\rm G$ & $0.05-3.0$ & $41370.151$ & $41336.591$ \\ \hline
\end{tabular}
\end{center}
\end{table*}
For the spatial distribution of stars with $R\ge1.0$~kpc, the bulge can be 
approximated with various kinds of density models \citep{dwek95} and the 
following three models are selected for the ML analysis: two Exponential types 
($\rm E_{1}$ and $\rm E_{2}$) and one Gaussian ($\rm G$) type which are given 
as follows \citep{dwek95,ratt07,cao13}: 
\begin{align}
\text{Model}~{\rm E_{1}}:\rho=&\rho_{0}\exp\left(-R\right) \\
\text{Model}~{\rm E_{2}}:\rho=&\rho_{0}k_{0}\left(R_{\rm s}\right) \\
\text{Model}~~{\rm G}:\rho=&\rho_{0}\exp\left({-0.5 R_{s}^{2}}\right).
\end{align}
Here $\rho_{0}$ is the unknown normalization constant  and is a function of 
$R,\theta,a,b,c$; $\theta$ is the angle at which the major axis lying
in the Galactic plane is tilted with respect to the line joining 
the GC and  the Sun; $a, b, c$ are the normalized axes of the ellipsoidal 
distribution of the bulge. Here $k_{0}$ denotes the modified Bessel 
function of second kind. Models $\rm E_{1},\rm E_{2}$ and $\rm G$ 
represent exponential, exponential boxy and Gaussian-triaxial bulges, 
respectively \citep{dwek95}. The expressions  of $R$ and 
$R_{s}$ are respectively given by \citep{dwek95}: 
\begin{align}
R=&\left[\left(\frac{x}{a}\right)^{2}+\left(\frac{y}{b}\right)^{2}+\left(\frac{z}{c}\right)^{2}\right]^\frac{1}{2}. \\
R_{\rm s}=&\left\{\left[\left(\frac{x}{a}\right)^{2}+\left(\frac{y}{b}\right)^{2}\right]^{2}+\left(\frac{z}{c}\right)^{4}\right\}^\frac{1}{4}.
\end{align}
The value of $\rho_{0}$ can be obtained using the normalization condition:
\begin{align}
\int_{V}\rho dV=& 1,
\end{align}
which can be expressed as 
\begin{align}
4\pi abc\rho_{0}\int_{R_{\rm min}}^{R_{\rm max}} R^{2} \rho dR=& 1.
\end{align}
The likelihood function for a given density model is defined as
\begin{align}
L=&\prod_{i=1}^n\rho_{i}(R;R_{0},\theta,a,b,c) \nonumber \\
\Rightarrow \ln{L}=&\sum\limits_{i=1}^n\ln{\rho_{i}(R;R_{0},\theta,a,b,c)},
\end{align}
where $n$ denotes the total number of stars. $R_{\rm max}$ and $R_{\rm min}$
denote the lower and the upper limits of the integration, respectively. 
We take $R_{\rm min}=0.05$~kpc and $R_{\rm max}=1.5,2.0, 2.5, 3.0, 3.5, 4.0$ 
kpc. From 
the Fig.~\ref{fig:gal_dist}, it can be easily seen that most of the stars lie 
within $R=4.0$~kpc and the number of stars falls rapidly above $R=5.0$~kpc. 
As logarithm is monotonically increasing function, we use log-likelihood 
rather than likelihood function for the mathematical convenience. We find the 
values of the parameters which maximizes $\ln{L}$ in five dimensional 
parameter space using python implemented limited-memory BFGS optimization 
method `L-BFGS' \citep{byrd95}. More refined values of these parameters and 
their statistical uncertainties are explored using MCMC 
analysis in a Bayesian framework \citep{fore13}. The posterior distribution 
is sampled with $1000$ walkers, $1000$ iterations and $100$ burn-in  using the 
affine-invariant MCMC ensemble sampler emcee-Python package \citep{fore13}.   

To find the ML parameters, we have provided initial guesses as well as the 
minimum and  the maximum limits for each of the  parameters. We have run 
`L-BFGS' several times based on different initial guesses within the range for 
convergence. The ML results obtained from the `L-BFGS' method are then used to 
initialize the walkers in a tiny Gaussian ball around these values for the 
MCMC analysis \citep{fore13}. The MCMC chains are used to compute the 
marginalized posterior distributions for each of the parameters. As the 
distributions are highly symmetrical as shown in Fig.~\ref{fig:emc}, the mean 
and the median of the distributions will be the same. The mean of each of the 
distribution ($50$th percentile) is taken as the robust measure of the best 
fit value with the statistical uncertainties measured up and down to the 
$16$th and $84$th percentiles, respectively. The results obtained from 
the MCMC analysis for different PL relations varying the intrinsic dispersion 
values are listed in Tables~\ref{table:tab34},~\ref{table:tab2},~
\ref{table:tab1},~\ref{table:tab07}. 
Tables~\ref{table:tab34},\ref{table:tab2},\ref{table:tab1} list the values 
of geometrical parameters employing \citet{ziaa19} PL relation considering 
$\sigma_{\rm in}=0.34,~0.2,~0.1$ mag, respectively. On the other hand, 
Table~\ref{table:tab07} lists the parameter values employing \citet{mcna11} 
PL relation with $\sigma_{\rm in}=0.07$ mag. A sampling acceptance rate 
of $\sim 0.5$ is achieved in each of the MCMC analyses.
\subsection{Best Fitting Model}
Bayesian Information Criterion(BIC) is a model selection criterion among a class
of parametric models \citep{schw78}. Akaike Information Criterion (AIC) is 
also a model selection criteria which measures the goodness of fit
of a statistical model \citep{akai74}. The best model minimizes both the AIC 
and BIC values for the given set of parameter values. They are defined 
as:
 \begin{align}
{\rm BIC}=&p\ln{n}-2\ln{\hat{L}}, \\
{\rm AIC}=&2p-2\ln{\hat{L}}.
\end{align}
Here $p$ denotes the number of parameters of the model; $n$, the total number 
of stars; $\hat{L}$, the maximum likelihood values given the parameters of the 
models. Among the three models considered in the present study, both the BIC 
and AIC values are the lowest for the `$\rm E_{1}$-model'. 
This implies that  $\rm E_{1}$-model is the best fitted model to the 
number density of $\delta$ Scuti stars for the \citet{ziaa19} PL relation 
assuming $\sigma_{\rm in}=0.1$ mag (Table~\ref{table:tab_bc}). However, this
is not the actual value of $\sigma_{\rm in}$.  The \citet{ziaa19} relation was 
obtained with $\sigma_{\rm in}=0.34$~mag. The AIC
and BIC values obtained using $\sigma_{\rm in}=0.34$~mag also suggest that the
model $\rm E_{1}$ is the best-fitted model among all the three models 
considered (Table~\ref{table:tab_bc}). Furthermore, comparing both the AIC and 
the BIC values obtained for the $\rm E_{1}$ model between \citet{ziaa19} with 
$\sigma_{\rm in}=0.34$ mag and \citet{mcna11}, it can be clearly seen from 
Table~\ref{table:tab_bc} that the use of \citet{ziaa19} relation with $\sigma_{\rm in}=0.34$ mag gives their lowest values. Therefore, the final values of the 
parameters as quoted in the present study are those obtained using the 
\citet{ziaa19} relation with $\sigma_{\rm in}=0.34$ mag. 
Fig.~\ref{fig:emc} depicts a representative 
plot of the marginalized posterior distribution and uncertainties of the model 
parameters obtained from the MCMC analysis for the $\rm E_{1}$-model up to 
$R_{\rm max}=3$~kpc conidering \citet{ziaa19} relation with 
$\sigma_{\rm in}=0.34$ mag. The statistical uncertainties are measured with 
the sample mean and $16$th (down) and $84$th (up) percentile differences. The 
black dashed vertical lines on each histogram represent the $16$th, $50$th and 
$84$th percentiles, respectively. 
\subsection{Systematic Uncertainty in $R_{0}$}
Systematic uncertainty in $R_{0}$ in the present study relies on two 
factors: errors in the extinction values as well as the errors due to the 
uncertainty in the coefficients of the PL relation $(\sim 0.090)$.
The error in $E(B-V)$ is calculated using $\sigma_{E(J-K_{s})}=0.02$~mag 
\citep{gonz12}. Taking into account all these errors, the uncertainty in 
distance modulus is found to be $\sigma_{\mu,{\rm sys}}=0.159$~mag which 
corresponds to a systematic uncertainty in $R_{0}$ with a value 
$\sigma_{R_{0},{\rm sys}}=0.586$~kpc.
\section{Summary and Conclusion}
\label{sec:summary}
In the present study, we have carried out an analysis of the structure of the
Galactic bulge using a clean sample of more than $7440$ high latitude 
$\delta$ Scuti stars ($\left|b\right|> 1^{\circ}$) recently released by the 
OGLE-IV survey. The distribution of the number density of stars as a function 
of different range of Galacto-centric distances has been modelled 
using three functions: two exponential types ($\rm E_{1}$ and 
$\rm E_{2}$) and one Gaussian type ($G$). The geometrical parameters of the 
bulge are calculated starting from $R_{\rm min}=0.05$~kpc up to various maximum 
values of $R_{\rm max}$ till $4$~kpc based on ML estimation. Refined values of 
these parameters along with their statistical uncertainties are explored using 
a fully Bayesian MCMC analysis. The values of the model parameters are found to 
vary slightly with different values of $R_{\rm max}$. Among all the three 
models chosen for the present study, it is found from the statistical analysis 
that the distribution of the number density of the bulge $\delta$ Scuti stars 
can be best fitted with the exponential triaxial model `{$\rm E_{1}$}' 
using the \citet{ziaa19} PL relation with $\sigma_{\rm in}=0.34$ mag.
It is to be noted that the other values of $\sigma_{\rm in}=0.1,~0.2$ mag were 
considered in the relation to see the effect of reduced values of 
$\sigma_{\rm in}$ and are not the real values. Therefore, the parameters corresponding to the  $\rm E_{1}$ model 
obtained with $\sigma_{\rm in }=0.34$ mag are chosen as the best parameters 
describing the structure of the bulge. The final values 
of the parameters are calculated from their weighted averages obtained with 
$R_{\rm max}=2.0,2.5,3.0,3.5,4.0$~kpc using the model $\rm E_{1}$ 
corresponding to the \citet{ziaa19} PL relation with $\sigma_{\rm in}=0.34$ 
mag. The errors of the parameters quoted for different values of $R_{\rm max}$ 
denote the statistical errors obtained from the MCMC analysis. The modelling 
of the bulge distribution with the best-fitted density model $\rm E_{1}$ yields 
the following results:
\begin{enumerate}
\item The mean distance to the GC is obtained as 
$R_{0}=8.034\pm0.012_{\rm stat}\pm0.586_{\rm sys}$~kpc which is consistent 
with most of the recent studies 
\citep{piet15,bhar17,grav19,griv19,griv20,griv21}.
\citet{piet15} estimated this value to be $8.27$~kpc using the distance 
distribution of the bulge RR Lyrae stars based on the OGLE-III data. Using 
a median statistical analysis of $28$ independent values of recent $R_{0}$, 
\citet{cama18} obtained $R_{0}=8.0\pm0.3$~kpc.  On the other hand, 
\citet{griv20,griv21} estimated the values of $R_{0}$ as $8.28\pm0.14$~kpc 
and $8.35\pm0.10$~kpc by employing a maximum likelihood analysis on the 
photometric data of OGLE RRab stars taken from \citet{sosz14} and  $715$ 
Type II Cepheids in the Galactic bulge  from VVV (VISTA Variables in the 
Via Lactea), respectively. Besides these values, \citet{grav19} found 
$R_{0}= 8.178\pm 0.013_{\rm stat}\pm0.022_{\rm sys}$~kpc following the star S2 
in its orbit around the massive black hole Sgr A$^{\star}$; this is a direct 
method without relying on any calibrators. By correcting the aberration effect 
from the same data as in \citet{grav19}, \citet{abut21} estimated 
$R_{0}=8.275\pm0.009_{\rm stat.}\pm0.033_{\rm sys.}$~kpc. However, 
\citet{do19} found $R_{0}=7.94\pm0.05_{\rm stat.}\pm0.03_{\rm sys.}$~kpc by 
independently analyzing the orbit of S2 star. On the other hand, using $21$ 
individual estimates of $R_{0}$ from $2017$ onwards, \citet{boby21b}  
calculated the best value to be $R_{0}=8.1\pm0.1$~kpc. 
\item The inclination angle made by the major axis lying in the Galactic 
plane with the Sun-GC line is found to be  
$\theta=22^{\circ}.006\pm2^{\circ}.078$. This value is quite in good agreement 
with other values as obtained in the literature (\citet{deka13,piet15,simi17,grad20,du20} and references therein). 
\item  The axes ratios are found to be 
$a:b:c=1.000\pm 0.005:0.348\pm0.002:0.421\pm0.002$. Smaller values of 
$b$ and $c$ as compared to the values of $a$ obtained for $R\ge 2.0$~kpc 
indicate that the distribution of the $\delta$ Scuti stars in the bulge 
follows a bar-like structure with a bar angle of 
$22^{\circ}.006\pm2^{\circ}.078$. 
\end{enumerate}
\section*{Acknowledgments} 
The author acknowledge the use of highly valuable publicly available OGLE-IV
photometric data of $\delta$ Scuti stars for this study. The authors thank 
the reviewer Prof.  Mart{\'\i}n {L{\'o}pez-Corredoira for useful comments 
and valuable suggestions which significantly improved the presentation of the 
manuscript.} Thanks are due to Prof. T. R. Bedding,  University of Sydney 
(Australia) for kindly informing us the value of intrinsic dispersion in the 
PL relation through an email. SD thanks Council of Scientific and Industrial 
Research (CSIR), Govt. of India, New Delhi for a financial support through the 
research grant ``03(1425)/18/EMR-II''. MD acknowledges CSIR for providing the 
Junior Research Fellowship (JRF) through CSIR-NET under the project. KK thanks 
CSIR for a senior research fellowship (SRF).  The paper makes use of the 
facility from \url{https://arxiv.org/archive/astro-ph}, NASA's Astrophysics Data
System (ADS) and SIMBAD data bases.
\section*{DATA AVAILABILITY}
The data underlying this article are available at 
\url{ftp://ftp.astrouw.edu.pl/ogle/ogle4/OCVS/blg/dsct/}. The derived data 
generated in this research will be shared on reasonable request to the
corresponding author.
\bibliographystyle{mnras}
\bibliography{ds_like}
\bsp
\label{lastpage}
\end{document}